\documentclass[aps,pra,reprint,groupedaddress,twocolumn]{revtex4-2}

\usepackage{graphicx}
\graphicspath{ {./images/} }
\usepackage{xcolor}
\usepackage{amsmath}
\usepackage{mathtools}
\usepackage{amssymb}
\usepackage{flushend}
\usepackage[T1]{fontenc}

\usepackage{braket}
\usepackage{amssymb}
\usepackage{amsmath}
\usepackage{natbib}
\usepackage{braket}
\usepackage{amssymb}
\usepackage{amsmath}
\usepackage{dsfont}
\usepackage{algorithm}
\usepackage{algpseudocode}

\newcommand{\SI}[1]{\ensuremath{\mathrm{#1}}}

\begin{document}

\title{A Throughput-Oriented Analytical Model for Post-Quantum Security Protocols}

\author{Ignazio Pedone}
\affiliation{nodeQ Limited, 71-75 Shelton Street, Covent Garden, London WC2H 9JQ, United Kingdom}
\author{Stefano Pirandola}
\affiliation{nodeQ Limited, 71-75 Shelton Street, Covent Garden, London WC2H 9JQ, United Kingdom}









\begin{abstract}
Growing awareness of the impact of quantum threat on classical cryptography directly translates into a growing demand for accurate network simulation tools capable of estimating the integration effects of quantum-safe cryptography in current systems. In particular, the adoption of Post-Quantum Cryptography (PQC) has a direct impact on the performance of network endpoints and transmission overhead. This also affects the scalability of widely adopted security protocols such as TLS and SSH. In this paper, we present a throughput-oriented analytical model that provides a tight upper bound on the maximum sustainable rate of post-quantum secure connection establishment in TLS and SSH. This model takes into account both endpoint and network capacity constraints, decomposing the handshake process into dominant cryptographic operation time and network transmission time. Identifying the bottleneck allows us to derive the achievable throughput in terms of handshakes per second. The experimental results provided show the accuracy of the model against the data obtained from an experimental testbed using, among others, NIST standard primitives from FIPS 203, 204, and 205, including ML-KEM and ML-DSA. Finally, we integrate our model into a network environment and demonstrate how it can be leveraged to enable efficient resource allocation among multiple endpoints, optimizing PQC traffic in multiple-unicast scenarios.
\end{abstract}



\maketitle

\section{Introduction}
\label{sec:intro}
Post-Quantum Cryptography (PQC) has emerged as an effective countermeasure to the advent of quantum computing and the resulting quantum threat. Although PQC relies on classical algorithms and primarily software-based implementations, the mathematical hard problems underlying these schemes introduce several challenges in terms of performance in real-world applications. For this reason, benchmarking, modelling, and simulation of these primitives in network scenarios are particularly important to evaluate, design, and support migration strategies from current classical cryptography to PQC systems.

Prior work \cite{Sikeridis2020PQTLSSSH,Sosnowski2023PQTLS13,Montenegro2026PQTLSFramework,Montenegro2026QUICvsTLS, Fitzgibbon2024ConstrainedBenchmarking} has analysed and discussed the performance of the new cryptographic primitives that emerged from the process of standardisation by the National Institute of Standards and Technology (NIST), their integration in security protocols such as TLS and SSH, and the overheads introduced in terms of computational complexity and data transmission over the network. In particular, these primitives have three desired characteristics that are quantifiable yet lead to practical trade-offs: quantum resistance, low computational cost, and small communication footprint. 

Quantum resistance is quantifiable according to the NIST security levels from one to five (higher is more secure) and is assessed under the Quantum Random Oracle Model (QROM), considering classical and quantum adversaries with finite computational resources. Computational costs and communication footprint depend on the family of primitives, i.e. the hard mathematical problem they are based on (e.g., lattice-based, code-based, hash-based), the specific algorithm implementation, and the NIST security level. A higher level of security for the same algorithm implies higher computational costs and a larger communication footprint (e.g., larger public key and signature sizes). Moreover, there are also non-quantifiable aspects such as the confidence in the quantum resistance for a specific family of algorithms with respect to others, depending on the maturity and years of study and analysis behind the cryptographic schemes. All these elements lead to security-performance trade-offs that justify the necessity of having accurate benchmarking for comparison in different scenarios. 

Another aspect is the computational and networking overheads that these new primitives introduce with respect to current asymmetric cryptography (e.g., RSA, ECDH, ECDSA). This is widely discussed in the studies above and largely covered by the scientific literature on the topic. Even though this consistent coverage exists, most of the work focuses on benchmarking over existing testbeds, comparison between existing security protocols integrating PQC, including open challenges and optimisation proposals, as well as new protocol variants. To the best of our knowledge, there is a lack of proposals for heuristic models that can simulate the behaviour of these primitives integrated in security protocols over realistic network scenarios, potentially on large-scale networks.

The idea behind this paper is to move in that direction and propose a simple analytical heuristic model capable of estimating the throughput of protocols such as TLS and SSH in terms of handshakes per second for secure connection establishment relying on the technical specifications of the node, the network capacity, and information about the adopted primitives. 

In particular, the contributions of this paper are: (i) a throughput-oriented decomposition of secure connection establishment into dominant asymmetric cryptographic operations and transmitted data budget, producing closed-form service rates for client, server, and network communication; (ii) a tight throughput upper bound designed for capacity planning, evaluating how many handshakes per second a pair of nodes over a TCP/IP network can sustain; (iii) a practical parametrisation for the model (e.g., based on CPU clock cycles, payload overheads, efficiency factors) which fits well with and allows to reuse existing benchmarks such as SUPERCOP~\cite{SUPERCOP}; (iv) an experimentally reproducible validation of the model against a real-world testbed to compare the accuracy of our model with the collected data, including both CPU-limited and bandwidth-limited regimes, identifying when non-cryptographic overhead and variance matter; (v) a full network integration of the model which translates handshake demand into bandwidth demand and vice versa, while optimally allocating resources between many competing endpoints in a multiple unicast scenario.

\section{Methodology}
\label{sec:methodology}
The starting point for describing our analytical model is to define the minimal operating scenario. The typical case is represented by two network endpoints that we refer to as client and server, with given technical specifications (Section~\ref{sec:gen_end_model}) and a direct link of fixed capacity (Section~\ref{sec:gen_net_model}) connecting them. This model also works with arbitrary endpoints. The choice of a client–server scenario simplifies the discussion of the selected security protocols, namely TLS and SSH. The objective is to calculate the TLS/SSH handshake rate (handshakes per second) for secure connection establishment under continuous load and steady-state operation. To this end, we model the secure connection establishment as a workload composed of independent handshake jobs that must be processed by three distinct system components: the client endpoint, the server endpoint, and the network connecting them. Each component is abstracted as a service system characterized by an aggregate service rate, expressed in completed handshakes per second. The sustainable handshake throughput of the overall system is then determined by the interaction of these service rates, and in particular by the slowest one among them (Section~\ref{sec:throughput_handshake_model}).

One modeling choice is to not consider round-trip time (RTT) or latency effects under the assumption of high concurrency, i.e., multiple handshakes running in parallel. Another assumption is that the handshake costs are dominated by asymmetric cryptographic operations and the transmission time for the PQC material. Other factors such as the residual protocol processing time, kernel and I/O operations, and packet loss are modeled as multiplicative overhead factors or additive delay terms, and they are generally environment-dependent, so they need to be estimated from measurements.

The core of the model consists in calculating the aggregate service rate for the various components. We start from the endpoints, which are identical in terms of the overall model, but differ between client and server, mainly because of the peculiarity of the specific security protocol. In the case of the endpoint model, to provide a useful parametrisation with accessible data to test diverse primitives, we used a decomposition based on CPU clock cycles to characterise the various cryptographic operations. Existing benchmarks, such as SUPERCOP, make this data available and allow the model to be used with various CPU architectures. For the network part, we analysed the data exchanged during the handshake of the specific protocols dividing them into fixed data, depending only on the protocol itself, and variable data that depend on the specific primitives used. Also in this case, data are accessible from the public submissions and specifications of the various PQC algorithms. 

After defining the model for TLS, we also extended it to SSH. This shows that the same model can be adapted to other security protocols reusing the same logic. We tested and validated our model using TLS against a testbed described in Section~\ref{sec:testbed_desc}. The first analysis consisted of two experiments: one in a regime where we used one logical CPU and limited resources on the client and server machines, and another where we used all the available resources to also understand how the model behaves in the case of multi-core architectures. In Section~\ref{sec:end_model_validation} we detail the primitives used and the results with a clear comparison between tested and simulated cases. The second analysis described in Section~\ref{sec:net_model_validation} focuses on a different perspective, where the bottleneck is the network and it is induced during the test through the throttling of the network interfaces.

In the final part of this work, we discuss the extension of the model to a network scenario, where we consider the problem of resource sharing at each node in case of multiple unicast connections. The goal is to allocate the node resources properly and rescale them according to fairness or user-defined policies. This operation has to be done at a low computational cost to avoid additional overhead in solving other network-related optimisation problems. We therefore demonstrate how our model enables efficient resource allocation and supports scalable network orchestration.

\section{General endpoint model}
\label{sec:gen_end_model}
This section describes the general endpoint model that is not tied to a particular security protocol. It defines a generic endpoint node and its technical specifications that are required to calculate the aggregate service rate or maximum handshake throughput sustainable by the endpoint. In particular, we define the technical specifications for the endpoint $e$ as:

\begin{itemize}
    \item $N_e$: \emph{Number of logical CPUs}.  
    This denotes the total number of schedulable CPU contexts available within a single shared-memory system (i.e., a single physical server or blade), as exposed to the operating system or hypervisor. Logical CPUs correspond to schedulable hardware threads, typically one per core when SMT is disabled and multiple per core when SMT is enabled, and may be distributed across multiple processor sockets and Non-Uniform Memory Access (NUMA) domains interconnected by intra-server interconnects (e.g., QPI). Logical CPUs share underlying execution and memory resources; therefore, parallel efficiency is not assumed to scale linearly with $N_e$. Cores belonging to distinct physical machines (e.g., separate blades connected via InfiniBand) are excluded and modeled as separate endpoints.

    \item $f_e \in \mathbb{R}_{>0}$: \emph{Clock frequency per logical CPU (Hz)}.  
    This is the nominal per-CPU clock rate used to convert CPU-cycle costs into execution times at endpoint~$e$. In practice, $f_e$ represents an effective average frequency over the measurement interval and may differ from the advertised base/boost frequencies due to dynamic frequency scaling and thermal limits.

    \item $\eta_e \in (0,1]$: \emph{CPU efficiency factor}.  
    This dimensionless factor accounts for the fact that not all nominal CPU cycles contribute to useful cryptographic computation. It absorbs microarchitectural and system-level effects such as pipeline stalls, cache and memory latency, instruction-mix inefficiency, virtualisation overhead and microarchitectural execution inefficiencies internal to a single logical CPU. Values closer to~$1$ correspond to near-ideal utilisation.

    \item $\alpha_e \in \mathbb{R}_{\ge 0}$: \emph{Non-cryptographic overhead fraction}.  
    This multiplicative overhead captures additional per-handshake CPU work not explicitly modeled as asymmetric cryptographic operations (e.g., protocol state processing, parsing/serialization, runtime management overhead, and user-space application logic). As shown in Eq.~\eqref{eq:endpoint_service_time}, $\alpha_e$ acts as a fractional increase relative to the modeled cryptographic time $T^{\mathrm{crypto}}_e$. It is environment- and implementation-dependent and is typically calibrated empirically.

    \item $\delta_e \in \mathbb{R}_{\ge 0}$: \emph{Additive I/O delay (s)}.  
    This term captures per-handshake delay components that are not proportional to cryptographic computation time and that may include blocking socket operations, connection setup and teardown, kernel-network stack latency, and context-switch overhead. In particular, $\delta_e$ models the effective waiting time during which handshake worker threads are stalled on I/O and therefore do not contribute to useful cryptographic computation. This phenomenon can manifest as reduced observed CPU utilization (e.g., 30\% utilization under I/O-bound operation). The value of $\delta_e$ is workload- and deployment-dependent and is calibrated empirically (e.g., by comparing \emph{Mode~I} and \emph{Mode~II} experiments in Section~\ref{sec:testbed_desc}).

    \item $\rho_e \in (0,1]$: \emph{Parallel efficiency factor}.
    The factor $\rho_e$ models the effective loss of parallel efficiency when executing across $N_e$ logical CPUs within a single shared-memory system. It captures the aggregate impact of non-ideal scaling effects, including resource sharing due to simultaneous multithreading (SMT), contention for shared caches and memory bandwidth, NUMA, and scheduling interference introduced by the operating system or hypervisor. As a result, $\rho_e$ reflects the deviation from ideal linear scaling and may depend on $N_e$ and the workload characteristics.
\end{itemize}

To add another building block to our model, we define the fundamental cryptographic operations related to public-key cryptography. As in the PQC standardisation process, we distinguish between two types of algorithms: Key Encapsulation Mechanism (KEM) and Digital Signature Schemes (SIG). The first enables key establishment, while the second allows performing digital signatures and verifying them. For both types, we have the key pair generation operation, which allows the creation of a pair of keys, a public and a private key, required for all the other operations. The KEM type also includes an encapsulation operation, which allows the sender to generate a ciphertext using the public key of the receiver, enabling the receiver to decapsulate it using its private key and derive a shared secret. The SIG type includes a signing operation, used to generate a digital signature using the sender's private key and a verification operation, used to validate the signature using the receiver's public key. More formally, let $\mathcal{O}$ denote the set of public-key cryptographic operations considered in the model, defined as $\mathcal{O} = \mathcal{O}_{\mathrm{KEM}} \;\cup\; \mathcal{O}_{\mathrm{SIG}}$, 
where $\mathcal{O}_{\mathrm{KEM}} = \{\mathrm{kg}_{\mathrm{kem}}, \mathrm{enc}, \mathrm{dec}\}$ and $\mathcal{O}_{\mathrm{SIG}} = \{\mathrm{kg}_{\mathrm{sig}}, \mathrm{sign}, \mathrm{verify}\}$.
For each operation $i \in \mathcal{O}$ and endpoint $e$, we define:
\begin{itemize}
  \item $c_{e,i} \in \mathbb{R}_{>0}$ as the number of CPU cycles required to execute
  operation $i$ on the specific CPU microarchitecture and cryptographic implementation used by endpoint $e$;
  \item $t_{e,i} \in \mathbb{R}_{>0}$ as the execution time of operation $i$ at
  endpoint $e$.
\end{itemize}

Let endpoint $e$ be characterized by a clock frequency $f_e$ and a CPU efficiency
factor $\eta_e$, then the execution time of operation $i$ at endpoint $e$ is given by
\begin{align}
t_{e,i} = \frac{c_{e,i}}{f_e \eta_e}.
\label{eq:time_operation}
\end{align}

We now introduce an intermediate quantity that captures the modeled
cryptographic workload performed by an endpoint during a single handshake,
independently of any specific security protocol.

Let $T^{\mathrm{crypto}}_e \in \mathbb{R}_{\ge 0}$ denote the
\emph{cryptographic computation time per handshake} at endpoint $e$.
This quantity represents the total execution time of the public-key
cryptographic operations required by a given security protocol instance,
assuming sequential execution of such operations at the endpoint. The specific
set of operations contributing to $T^{\mathrm{crypto}}_e$ is
protocol-dependent and will be instantiated in later sections (e.g., for TLS
or SSH), but the abstraction itself is protocol-agnostic.

Introducing $T^{\mathrm{crypto}}_e$ provides a clean separation between
(i) the cryptographic workload imposed by a security protocol and
(ii) the endpoint-specific factors that determine how this workload translates
into an effective service time and throughput. This separation enables the same
endpoint model to be reused across different security protocols and
cryptographic configurations by simply redefining $T^{\mathrm{crypto}}_e$.

Non-cryptographic processing performed by the endpoint during a handshake is
modeled via the multiplicative overhead factor $\alpha_e$, while fixed
per-handshake delays not proportional to computation (e.g., I/O and kernel
effects) are captured by the additive term $\delta_e$. The resulting
\emph{endpoint service time} is therefore defined as
\begin{align}
T_e
=
(1+\alpha_e)\,T^{\mathrm{crypto}}_e
+
\delta_e.
\label{eq:endpoint_service_time}
\end{align}

Finally, handshake processing can be parallelized across the $N_e$ logical CPUs
available at endpoint $e$, with non-ideal scaling captured by the parallel
efficiency factor $\rho_e \in (0,1]$. The corresponding \emph{endpoint service
rate}, expressed in completed handshakes per second, is given by
\begin{align}
\mu_e = \frac{\rho_e N_e}{T_e}.
\label{eq:endpoint_service_rate}
\end{align}

\section{General Network Model}
\label{sec:gen_net_model}

This section focuses on describing the network model, which in our case is based on a generic capacity-limited classical communication channel responsible for the transmission of a finite amount of data associated with the establishment or configuration of a secure communication between two endpoints. In particular, we are interested in modeling the volume of data exchanged during the handshake phases of a security protocol and more specifically the ones related to public-key cryptography such as key establishment and authentication. We avoid considering the sustained application data transfer.

As mentioned in Section~\ref{sec:methodology}, the model is throughput-oriented and intentionally abstracts away latency and RTT effects under the high-concurrency assumption. The objective is to focus on the average service time required to transmit a given amount of data over the channel. This choice is motivated by the assumption that in highly parallel and short-lived connection scenarios, the long-term performance is dominated by service rates rather than by individual exchange latency. The network model is formally defined as:

\begin{itemize}
  \item $C \in \mathbb{R}_{>0}$: \emph{channel capacity in bits per second (bps)};
  \item $p \in [0,1)$: \emph{packet loss probability};
  \item $B \in \mathbb{R}_{>0}$: \emph{number of bytes} that must be transmitted during the handshake phases.
\end{itemize}

Packet loss is modeled as an effective increase in transmitted data volume due to retransmissions. The expected number of bytes transmitted over the channel is therefore
\begin{equation}
   B_{\mathrm{tx}} = \frac{B}{1 - p}. 
\end{equation}
The network service time associated with transmitting $B_{\mathrm{tx}}$ bytes
over a channel of capacity $C$ is given by
\begin{align}
T_{\mathrm{net}} = \frac{8 \, B_{\mathrm{tx}}}{C} = \frac{8 \, B}{C (1 - p)}.
\end{align}
The corresponding network service rate is
\begin{align}
\mu_{\mathrm{net}} = \frac{1}{T_{\mathrm{net}}} = \frac{C (1 - p)}{8 \, B}.
\label{eq:mu_net}
\end{align}

This service time formulation allows the network component to be treated on equal footing with endpoint-side service times, enabling direct comparison when identifying throughput bottlenecks in the composite protocol model.

\section{Security protocol composite model}
\label{sec:throughput_handshake_model}
For our throughput-oriented composite model we adopt for convenience a client-server approach, but as mentioned before in Section~\ref{sec:methodology} this can be applied to a generic unicast connection between two endpoints. The model involves multiple processing stages executed by distinct components, namely the client endpoint, the server endpoint, and the communication channel. Each component is treated as an independent service system characterized by an average service rate, measured in completed handshakes per second. For systems composed of multiple
service stages arranged in series, the steady-state throughput is upper bounded by the minimum service rate among the individual stages, independently of per-request latency.

\medskip
Let $\mu_{\mathrm{client}}$, $\mu_{\mathrm{server}}$, and $\mu_{\mathrm{net}}$ denote the service rates associated with the client endpoint, server endpoint, and network or communication channel, respectively. The overall handshake throughput is therefore bounded by
\begin{align}
\mu_{\mathrm{unicast}}
=
\min \left(
\mu_{\mathrm{client}},
\mu_{\mathrm{server}},
\mu_{\mathrm{net}}
\right).
\end{align}

The corresponding bottleneck-implied service time for a handshake is
\begin{align}
T_{\mathrm{unicast}}
=
\max \left(
\mu_{\mathrm{client}}^{-1},
\mu_{\mathrm{server}}^{-1},
\mu_{\mathrm{net}}^{-1}
\right).
\end{align}

\subsection{TLS protocol}
\label{sec:tls_model}
This section describes in detail the application of our model to TLS 1.3 and in particular it references the TLS handshake steps and message flow defined in~\cite{Montenegro2026PQTLSFramework} and based on RFC 8446 \cite{RFC8446} and more recent IETF drafts \cite{TLSHybridDesign12}. First, we assume that we only have server authentication so all the steps related to the potential client authentication such as in mTLS case can be skipped. Second, we do not model the PKI, in the sense that we do not consider the intermediate CAs and the other potential certificate verification mechanisms. We only rely on the verification of the signature performed by the server against a public key trusted by the client. These assumptions simplify the model but can be added to the calculation if needed as we describe in the following.

A full TLS 1.3 handshake on the endpoints side, according to these assumptions, requires the client to perform a KEM key generation, a KEM decapsulation, and a SIG verify. On the server side the operations are only a KEM encapsulation and SIG sign. For a \emph{single handshake}, the cryptographic operations executed at each endpoint are sequentially ordered by the protocol logic. However, under steady load and high concurrency, operations belonging to \emph{different handshakes} may be arbitrarily interleaved and executed in parallel across available CPU resources. This observation justifies the throughput-oriented abstraction adopted in this work: we model the per-handshake cost assuming sequential execution, while aggregate throughput is obtained via parallel service rates.

Let $T^{\mathrm{crypto}}_e$ denote the total cryptographic computation time per handshake at endpoint $e$ as
\begin{equation}
T^{\mathrm{crypto}}_e
=
\sum_{i \in \mathcal{O}_e} t_{e,i},
\label{eq:Tcrypto_sum}
\end{equation}
where $\mathcal{O}_e$ is the set of operations associated with endpoint $e$.

Following this, and as we have already mentioned, on the \emph{client} side, the handshake requires:
(i) generation of an ephemeral KEM key pair for the \texttt{ClientHello},
(ii) decapsulation of the server’s KEM ciphertext, and
(iii) verification of the server’s authentication signature. The corresponding cryptographic time is therefore
\begin{equation}
T^{\mathrm{crypto}}_{\mathrm{client}}
=
t_{\mathrm{kg_{kem}}}
+
t_{\mathrm{dec}}
+
t_{\mathrm{verify}} .
\label{eq:tls_Tcrypto_client}
\end{equation}

On the \emph{server} side, the handshake requires:
(i) encapsulation to the client’s KEM public key and
(ii) generation of a digital signature over the handshake transcript.
The corresponding cryptographic time is
\begin{equation}
T^{\mathrm{crypto}}_{\mathrm{server}}
=
t_{\mathrm{enc}}
+
t_{\mathrm{sign}} .
\label{eq:tls_Tcrypto_server}
\end{equation}
Additional certificate verification for intermediate CAs can be included in the model by adding on the client side the needed verifications as a sum over the number of the intermediate certificates. For mutual authentication, on the client side, we would need an additional signature operation, while, on the server side, a verification operation, or more if we consider all the intermediate CAs signatures. This is again out of the scope of this work. All per-operation execution times are computed using Eq.~\eqref{eq:time_operation}. Endpoint service times then follow from Eq.~\eqref{eq:endpoint_service_time}. Incorporating non-cryptographic processing and fixed per-handshake delays, the endpoint service times are:
\begin{align}
T_{\mathrm{client}}
&=
(1+\alpha_{\mathrm{client}})\,T^{\mathrm{crypto}}_{\mathrm{client}}
+
\delta_{\mathrm{client}},
\label{eq:Tclient}
\\
T_{\mathrm{server}}
&=
(1+\alpha_{\mathrm{server}})\,T^{\mathrm{crypto}}_{\mathrm{server}}
+
\delta_{\mathrm{server}}.
\label{eq:Tserver}
\end{align}

Substituting into the endpoint service-rate definition Eq.~\eqref{eq:endpoint_service_rate} we obtain:
\begin{align}
\mu_{\mathrm{client}}
&=
\frac{\rho_{\mathrm{client}}\,N_{\mathrm{client}}}
{(1+\alpha_{\mathrm{client}})\,T^{\mathrm{crypto}}_{\mathrm{client}}+\delta_{\mathrm{client}}},
\label{eq:mu_client_subst}
\\
\mu_{\mathrm{server}}
&=
\frac{\rho_{\mathrm{server}}\,N_{\mathrm{server}}}
{(1+\alpha_{\mathrm{server}})\,T^{\mathrm{crypto}}_{\mathrm{server}}+\delta_{\mathrm{server}}}.
\label{eq:mu_server_subst}
\end{align}

Under the \emph{infinite-bandwidth assumption}, i.e., when network transmission
does not constitute a bottleneck, the sustainable TLS handshake throughput is
entirely endpoint-limited and is given by
\begin{equation}
\mu_{\mathrm{TLS}}^{\mathrm{CPU}}
=
\min\!\left(
\mu_{\mathrm{client}},
\mu_{\mathrm{server}}
\right).
\label{eq:tls_cpu_limited}
\end{equation}

When the network is not negligible, the TLS handshake throughput is also constrained
by the volume of data exchanged during connection establishment.
For a full TLS~1.3 handshake with server authentication and without \texttt{HelloRetryRequest}, the handshake messages involved, as specified in RFC~8446, are:
\begin{itemize}
    \item \texttt{ClientHello};
    \item \texttt{ServerHello};
    \item \texttt{EncryptedExtensions};
    \item \texttt{Certificate};
    \item \texttt{CertificateVerify};
    \item \texttt{Finished} messages (one sent by the server and one by the client).
\end{itemize}

These messages convey all information required for secure connection establishment,
including the negotiated TLS protocol version, supported groups and cipher suites,
authentication material (server certificate and certificate chain),
supported signature algorithms, and additional protocol extensions.
From a throughput-oriented perspective, however, not all transmitted data
contributes equally to the network service time. In our model, we therefore decompose the total on-the-wire handshake traffic into a \emph{baseline} component and a \emph{primitive-dependent} component.

The baseline component, denoted by $B_{\mathrm{fix}}$, captures protocol framing
(e.g., TLS record and handshake headers, fixed fields, extension framing, cipher-suite
negotiation) and any deployment-specific but \emph{primitive-independent} payload
that remains constant across the cryptographic configurations under comparison
(e.g., a fixed certificate chain encoding and length).
In practice, $B_{\mathrm{fix}}$ is treated as a calibrated constant obtained from
a reference handshake trace for the considered implementation and configuration.

The primitive-dependent component captures the dominant algorithm-dependent
contribution introduced by public-key cryptography. We denote this component by
$B_{\mathrm{var}}$ and define it as
\begin{equation}
B_{\mathrm{var}} = b_{\mathrm{pk}} + b_{\mathrm{ct}} + b_{\mathrm{cer}} + b_{\mathrm{sig}} .
\end{equation}

The individual terms are defined as follows:
\begin{itemize}
    \item $b_{\mathrm{pk}}$: bytes carried in the \texttt{ClientHello} message within
    the \texttt{key\_share} extension, corresponding to the post-quantum KEM public
    key generated by the client;

    \item $b_{\mathrm{ct}}$: bytes carried in the \texttt{ServerHello} message within
    the \texttt{key\_share} extension, corresponding to the post-quantum KEM
    ciphertext generated by the server in response to the client key share;

    \item $b_{\mathrm{cer}}$: primitive-dependent bytes that are carried in the \texttt{Certificate} message, including the public key and any additional fields whose size depends on the selected authentication algorithm (e.g., post-quantum public key material and associated algorithm-specific encodings). Any remaining certificate-chain payload that is primitive-independent is accounted for in $B_{\mathrm{fix}}$;

    \item $b_{\mathrm{sig}}$: bytes carried in the \texttt{CertificateVerify} message,
    corresponding to the digital signature generated by the server using the
    selected authentication algorithm over the hash of the TLS handshake transcript.
\end{itemize}

With this decomposition, the total number of bytes transmitted during a TLS~1.3
handshake is given by
\begin{equation}
    B_{\mathrm{TLS}} = B_{\mathrm{fix}} + B_{\mathrm{var}} .
\end{equation}

Using the general network model introduced in
Section~\ref{sec:gen_net_model}, the resulting network service rate is
\begin{equation}
\mu_{\mathrm{net}}
=
\frac{C(1-p)}{8\,B_{\mathrm{TLS}}}.
\label{eq:tls_net_rate}
\end{equation}

When the network service time is not negligible, endpoint-side and network-side
constraints must be considered jointly. Applying the composite throughput rule
introduced in Section~\ref{sec:throughput_handshake_model}, the sustainable
TLS~1.3 handshake throughput is therefore
\begin{equation}
\mu_{\mathrm{TLS}}
=
\min\!\left(
\mu_{\mathrm{client}},
\mu_{\mathrm{server}},
\mu_{\mathrm{net}}
\right).
\label{eq:tls_total_rate}
\end{equation}

This expression explicitly identifies whether the achievable handshake rate is
limited by client-side computation, server-side computation, or by the transmission
of post-quantum handshake material over the network.

\subsection{SSH protocol}
\label{sec:ssh_model}

The throughput-oriented framework introduced for TLS can be extended to the
Secure Shell protocol (SSHv2) by instantiating the endpoint cryptographic workload
and the handshake byte budget according to the SSH connection establishment
procedure. As for TLS, the general endpoint model, network model, and composite
bottleneck rule introduced in
Sections~\ref{sec:gen_end_model}, \ref{sec:gen_net_model}, and
\ref{sec:throughput_handshake_model} are reused without modification.

In SSHv2, secure channel establishment is composed of multiple logical phases,
namely the transport-layer protocol (including key exchange), followed by server
authentication and, optionally, client authentication. In this work, we follow
the post-quantum SSH handshake structure introduced and experimentally evaluated
in~\cite{Sikeridis2020PQTLSSSH}, which reflects post-quantum–enabled OpenSSH
implementations based on \texttt{liboqs}.

\medskip
\noindent
In the post-quantum SSH variant considered, the transport-layer key exchange is
realized as a KEM exchange using the \texttt{SSH\_MSG\_KEXPQ\_INIT} and \texttt{SSH\_MSG\_KEXPQ\_REPLY} message pair. The client generates an ephemeral post-quantum KEM key pair and sends the public key to the server, which responds with a corresponding post-quantum ciphertext. The server authenticates the exchange by signing the exchange hash $H$, which the client verifies. When public-key user authentication is enabled, the client subsequently signs an authentication transcript, which the server verifies.

As for TLS, cryptographic operations belonging to a single SSH handshake are
sequentially ordered by the protocol logic. Under steady-state operation and high
concurrency, however, cryptographic operations belonging to different handshakes
may be interleaved and executed in parallel across available CPU resources. This
justifies modeling the per-handshake cryptographic cost as a sequential workload,
while aggregate performance is captured through endpoint service rates.

Let $T^{\mathrm{crypto}}_e$ denote the total cryptographic computation time per
handshake at endpoint $e$, defined as the sum of the execution times of the
public-key operations required by the protocol.

On the \emph{client} side, the SSH handshake requires:
(i) generation of an ephemeral post-quantum KEM key pair,
(ii) decapsulation of the server’s post-quantum ciphertext, and
(iii) verification of the server’s host-key signature over the exchange hash $H$.
If public-key user authentication is enabled, the client additionally performs a
signature operation over the authentication transcript. We model the presence of public-key user authentication through a binary
indicator variable $\chi_{\mathrm{auth}} \in \{0,1\}$, where
$\chi_{\mathrm{auth}} = 1$ denotes that client public-key authentication is
enabled, and $\chi_{\mathrm{auth}} = 0$ otherwise. The resulting
cryptographic time is therefore
\begin{equation}
T^{\mathrm{crypto}}_{\mathrm{client}}
=
t_{\mathrm{kg_{kem}}}
+
t_{\mathrm{dec}}
+
t_{\mathrm{verify}}
+
\chi_{\mathrm{auth}}\, t_{\mathrm{sign}} .
\end{equation}

On the \emph{server} side, the handshake requires:
(i) encapsulation to the client’s post-quantum KEM public key and
(ii) generation of a digital signature over the exchange hash $H$ using the
server host key. If public-key user authentication is enabled, the server
additionally verifies the client’s authentication signature. The corresponding
cryptographic time is
\begin{equation}
T^{\mathrm{crypto}}_{\mathrm{server}}
=
t_{\mathrm{enc}}
+
t_{\mathrm{sign}}
+
\chi_{\mathrm{auth}}\, t_{\mathrm{verify}} .
\end{equation}

Endpoint service times and service rates are then obtained from the general
endpoint model by including non-cryptographic overhead and fixed per-handshake
delays, yielding $\mu_{\mathrm{client}}$ and $\mu_{\mathrm{server}}$ as in
Section~\ref{sec:tls_model}.

\medskip
\noindent
When the network is not negligible, the SSH handshake throughput is also
constrained by the volume of data exchanged during connection establishment.
The SSH handshake messages involved include the initial protocol identification
exchange, algorithm negotiation messages, post-quantum key exchange messages,
server authentication data, and, optionally, client authentication data.

As for TLS, we decompose the total on-the-wire SSH handshake traffic into a fixed
component and a variable, algorithm-dependent component. The fixed component
includes protocol identification strings, framing, and algorithm negotiation
lists, whose size does not significantly depend on the selected cryptographic
primitives.

The variable component captures the dominant post-quantum contribution and is
defined as
\begin{equation}
B_{\mathrm{var}}^{\mathrm{SSH}}
=
b_{\mathrm{pk}}
+
b_{\mathrm{ct}}
+
b_{\mathrm{sig}}^{\mathrm{srv}}
+
\chi_{\mathrm{auth}}
\left(
b_{\mathrm{pk}}^{\mathrm{cli}}
+
b_{\mathrm{sig}}^{\mathrm{cli}}
\right),
\end{equation}
where:
\begin{itemize}
  \item $b_{\mathrm{pk}}$ denotes the client post-quantum KEM public key carried in
  \texttt{SSH\_MSG\_KEXPQ\_INIT};
  \item $b_{\mathrm{ct}}$ denotes the post-quantum KEM ciphertext carried in
  \texttt{SSH\_MSG\_KEXPQ\_REPLY};
  \item $b_{\mathrm{sig}}^{\mathrm{srv}}$ denotes the server host-key signature over
  the exchange hash $H$;
  \item $b_{\mathrm{pk}}^{\mathrm{cli}}$ and $b_{\mathrm{sig}}^{\mathrm{cli}}$ denote,
  respectively, the client public key and the client authentication signature
  transmitted during public-key user authentication.
\end{itemize}

The total number of bytes transmitted during SSH connection establishment is then
\begin{equation}
B_{\mathrm{SSH}}
=
B_{\mathrm{fix}}^{\mathrm{SSH}}
+
B_{\mathrm{var}}^{\mathrm{SSH}} .
\end{equation}

Using the general network model of Section~\ref{sec:gen_net_model}, the resulting
network service rate is
\begin{equation}
\mu_{\mathrm{net}}
=
\frac{C(1-p)}{8\,B_{\mathrm{SSH}}}.
\end{equation}

Combining endpoint and network constraints, the sustainable SSH handshake
throughput under continuous load is therefore given by
\begin{equation}
\mu_{\mathrm{SSH}}
=
\min\!\left(
\mu_{\mathrm{client}},
\mu_{\mathrm{server}},
\mu_{\mathrm{net}}
\right).
\label{eq:ssh_total_rate}
\end{equation}

\section{Validation}
\label{sec:validation}

\subsection{Testbed Description}
\label{sec:testbed_desc}
\begin{figure}[ht]
    \centering
    \includegraphics[width=0.50\textwidth]{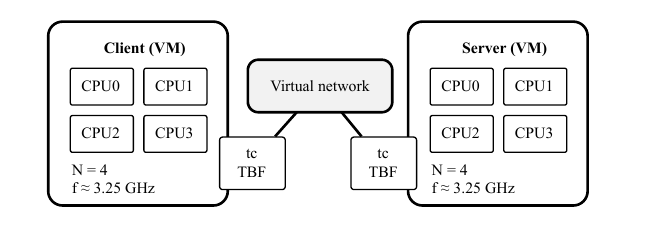}
    \caption{Design of the testbed to collect experimental data.}
    \label{fig:testbed}
\end{figure}
For model testing and validation, we collected experimental data from the testbed presented in Fig.~\ref{fig:testbed}. We compare the experimental data with the results generated by our simulation model. In particular, the testbed consists of two virtual machine instances (VMs) on Microsoft Azure of type \texttt{D4as\_v5}, each equipped with 4~vCPUs, 16~GB of RAM, and a 128~GB Premium SSD (LRS). The instances were interconnected through a virtual network providing up to 5~Gbps of bandwidth. Both machines run Ubuntu 24.04 LTS. The physical host CPU model was an AMD EPYC~7763 processor (64~cores, 128~threads), operating at 2.45~GHz with a boost frequency up to 3.5~GHz.

For validation of the TLS endpoint model, we reimplemented the TLS handshake using the \texttt{liboqs} library. Specifically, we developed a client--server utility deployed on both endpoints, allowing configuration of the number of client and server workers as well as the number of threads allocated to serve TLS handshake requests. The implementation follows the model described in Section~\ref{sec:throughput_handshake_model} and the relevant standards. The only simplification concerns the dominant components, namely the cryptographic operations and the data exchange. The results are consistent with those obtained for the composite model using the official OpenSSL~3.5.4 library~\cite{OpenSSL35}, as discussed below.

The experimental methodology required selecting between two operational modes: (i) \emph{Mode~I}, where a single TCP connection is established per TLS handshake, and (ii) \emph{Mode~II}, where multiple TLS handshakes are performed over a single TCP connection. Standard OpenSSL benchmarking tools typically operate in \emph{Mode~I}; therefore, we implemented \emph{Mode~II} in our utility, which also provides flexible resource allocation.

For the composite model, focusing primarily on networking bottlenecks, we directly relied on the OpenSSL~3.5 implementation using the \texttt{alpine/openssl:3.5.4} Docker image available on Docker Hub. This provides a ready-to-use OpenSSL~3.5.4 library with the included testing tools such as \texttt{openssl s\_time} and \texttt{openssl s\_connect}. In addition, we used Linux traffic control (\texttt{tc}) with a token bucket filter (TBF) to rate-limit the Docker container network interface and analyze handshake throughput under bandwidth-constrained conditions.

In the following, we focus exclusively on the TLS protocol model. The SSH model is left as an extension to demonstrate its flexibility and as potential future work for comparing performance across different security protocols.  

\subsection{Endpoint model validation}
\label{sec:end_model_validation}

For the endpoint model validation, we conducted two experimental test scenarios, each
evaluated both on the virtual testbed and on the proposed analytical model
implemented via software. In all tests, the network capacity was intentionally
left unconstrained so that the network service time was negligible. This
eliminates any bandwidth-induced bottleneck and reduces the problem to
identifying the computational bottleneck between the client and server
endpoints, i.e., the CPU-limited regime described in Section~\ref{sec:gen_end_model}.

The first test scenario allocates one logical CPU per endpoint with $8$ parallel
worker threads and an estimated parallel efficiency factor $\rho_{\mathrm{client}}=\rho_{\mathrm{server}}=1.0$.
Multiple TLS handshakes are multiplexed over the same TCP connection (\emph{Mode~II} in
the testbed description), so that the additive I/O delay terms
$\delta_{\mathrm{client}}$ and $\delta_{\mathrm{server}}$ are effectively
negligible and are therefore set to zero in the model.

The second test scenario follows the same logic but increases the computational
parallelism by allocating $4$ logical CPUs and $32$ worker threads per endpoint,
with an estimated parallel efficiency factor $\rho_{\mathrm{client}}=\rho_{\mathrm{server}}=0.5$, reflecting non-ideal
scaling at higher concurrency.

\begin{figure}[ht]
    \centering
    \includegraphics[width=0.48\textwidth]{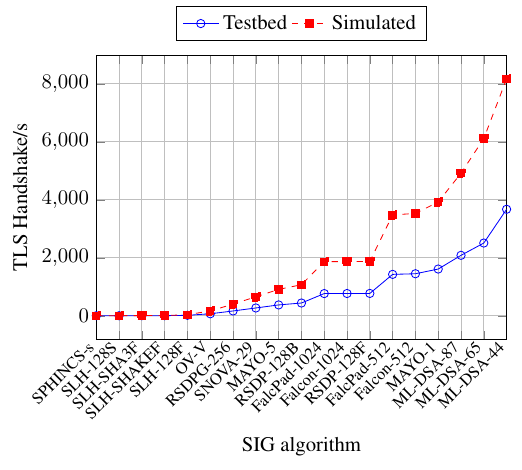}
    \caption{TLS single-CPU throughput with ML-KEM-1024.}
    \label{fig:ml-kem-exp1}
\end{figure}

\begin{figure}[ht]
    \centering
    \includegraphics[width=0.48\textwidth]{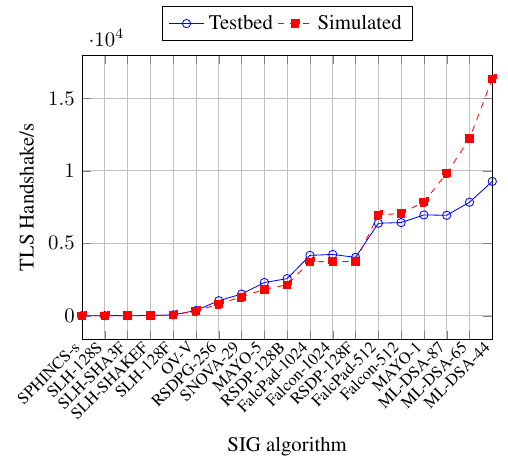}
    \caption{TLS multi-CPU throughput with ML-KEM-1024.}
\label{fig:ml-kem-exp2}
\end{figure}

\begin{figure}[ht]
    \centering
    \includegraphics[width=0.48\textwidth]{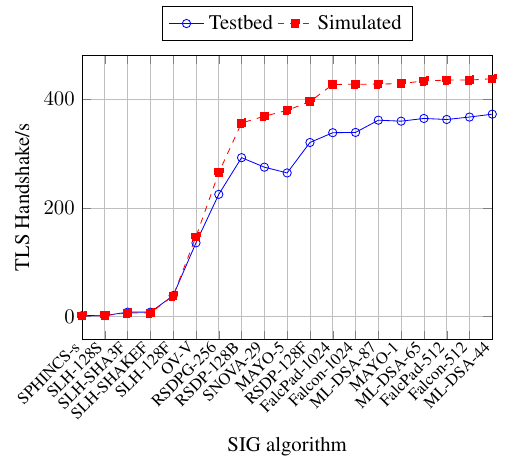}
    \caption{TLS single-CPU throughput with FRODO-KEM-1344-AES.}
    \label{fig:frodo-kem-exp1}
\end{figure}

\begin{figure}[ht]
    \centering
    \includegraphics[width=0.48\textwidth]{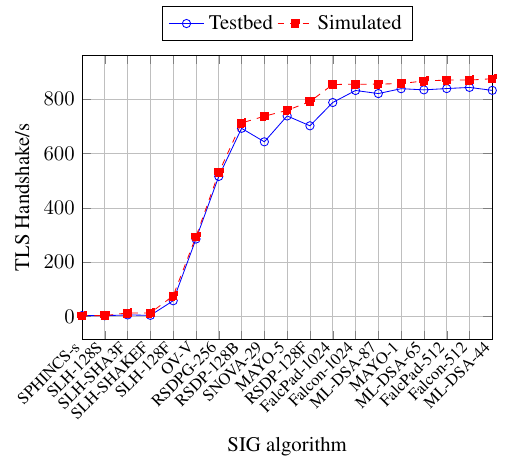}
    \caption{TLS multi-CPU throughput with FRODO-KEM-1344-AES.}
\label{fig:frodo-kem-exp2}
\end{figure}

Both test scenarios were executed under two cryptographic configurations. In the
first configuration, ML-KEM-1024 was used as the KEM while the signature algorithm
was varied across different post-quantum families. In the second configuration,
Frodo-KEM-1344-AES was used as the KEM with the same variation of signature
primitives. The ML-KEM-1024 results are reported in Fig.~\ref{fig:ml-kem-exp1} and Fig.~\ref{fig:ml-kem-exp2}, while the Frodo-KEM-1344-AES results are shown in Fig.~\ref{fig:frodo-kem-exp1} and Fig.~\ref{fig:frodo-kem-exp2}.

Figure~\ref{fig:ml-kem-exp1} shows that, for ML-KEM-1024, the simulated throughput closely follows the same trend observed on the testbed, confirming that the
handshake performance is dominated by asymmetric cryptographic computation, as
predicted by the endpoint model. However, ML-KEM-1024 is not among the most
computationally intensive KEMs. When paired with relatively lightweight
signature primitives (e.g., Falcon-1024), the gap between measured and simulated
results increases. In this experiment, both client and server are configured
with $N_{\mathrm{client}}=N_{\mathrm{server}}=1$ logical CPU operating at
$f_{\mathrm{client}}=f_{\mathrm{server}}=3.25\,\mathrm{GHz}$ and CPU efficiency
$\eta_{\mathrm{client}}=\eta_{\mathrm{server}}=0.75$. The non-cryptographic
overhead fractions are set to $\alpha_{\mathrm{client}}=0.54$ and
$\alpha_{\mathrm{server}}=0.10$. Parallel processing uses 8 worker threads with
parallel efficiency $\rho_{\mathrm{client}}=\rho_{\mathrm{server}}=1.0$. The
plotted curves compare throughput obtained from the analytical model with
throughput measured on the experimental testbed.

An additional effect is visible for the ML-DSA variants appearing in the upper
range of the plot. In this case, the CPU-cycle measurements collected on the
testbed and compared against SUPERCOP exhibit a comparatively large standard
deviation, leading to reduced stability and higher uncertainty in the estimated
cycle counts. In the current simulation, the non-cryptographic CPU overhead
fraction $\alpha_e$ is averaged across all primitive combinations, which further
amplifies this mismatch for primitives with higher intrinsic variability. A more
accurate calibration of per-primitive cycle counts and a refined modeling of
$\alpha_e$ would mitigate this effect and improve prediction accuracy for these
cases.

In Figure~\ref{fig:ml-kem-exp2}, increased parallelism reduces the discrepancy
observed for lightweight signature primitives, although the impact of CPU-cycle
variance remains visible. Overall, the monotonic trend and the predicted upper
bound are preserved, confirming the validity of the throughput-oriented model.
Nevertheless, tighter bounds for a specific architecture require more precise
cycle characterization and architecture-aware benchmarking. In this experiment,
both client and server are configured with
$N_{\mathrm{client}}=N_{\mathrm{server}}=4$ logical CPUs operating at
$f_{\mathrm{client}}=f_{\mathrm{server}}=3.25\,\mathrm{GHz}$ and CPU efficiency
$\eta_{\mathrm{client}}=\eta_{\mathrm{server}}=0.75$. The non-cryptographic
overhead fractions are set to $\alpha_{\mathrm{client}}=0.54$ and
$\alpha_{\mathrm{server}}=0.10$. Parallel processing uses 32 worker threads with
parallel efficiency $\rho_{\mathrm{client}}=\rho_{\mathrm{server}}=0.5$.

When moving to Frodo-KEM-1344-AES, Figure~\ref{fig:frodo-kem-exp1} shows a tighter
agreement between simulation and testbed measurements. This behavior is expected
because Frodo-based KEMs impose substantially higher computational costs, making
the cryptographic workload strongly dominant with respect to auxiliary overheads
and lightweight signature variations. This result provides strong validation of
the endpoint model in regimes where asymmetric cryptography clearly dominates
the processing pipeline. This behavior differs from the network-limited regime,
which is analyzed separately in the following. In this experiment, both client
and server are configured with
$N_{\mathrm{client}}=N_{\mathrm{server}}=1$ logical CPU operating at
$f_{\mathrm{client}}=f_{\mathrm{server}}=3.25\,\mathrm{GHz}$ and CPU efficiency
$\eta_{\mathrm{client}}=\eta_{\mathrm{server}}=0.75$. The non-cryptographic
overhead fractions are set to $\alpha_{\mathrm{client}}=0$ and
$\alpha_{\mathrm{server}}=0$. Parallel processing uses 8 worker threads with
parallel efficiency $\rho_{\mathrm{client}}=\rho_{\mathrm{server}}=1.0$.

Finally, Fig.~\ref{fig:frodo-kem-exp2} confirms that the increased parallelism further
improves the accuracy of the model also in the Frodo-KEM-1344-AES configuration.
In this case, the predicted bound is particularly tight, indicating that the
parallel efficiency factor $\rho_e$ and the aggregate CPU service model provide a
reliable approximation of the effective endpoint throughput in highly
compute-intensive scenarios. In this experiment, both client and server are
configured with $N_{\mathrm{client}}=N_{\mathrm{server}}=4$ logical CPU operating at
$f_{\mathrm{client}}=f_{\mathrm{server}}=3.25\,\mathrm{GHz}$ and CPU efficiency
$\eta_{\mathrm{client}}=\eta_{\mathrm{server}}=0.75$. The non-cryptographic
overhead fractions are set to $\alpha_{\mathrm{client}}=0$ and
$\alpha_{\mathrm{server}}=0$. Parallel processing uses 32 worker threads with
parallel efficiency $\rho_{\mathrm{client}}=\rho_{\mathrm{server}}=0.5$.

\subsection{Network model validation}
\label{sec:net_model_validation}

To validate the network model, we used the same Azure-based testbed described above. In particular, two Docker containers were deployed: one acting as TLS client and the other as TLS server, each running on a dedicated VM instance. Both containers executed OpenSSL version~3.5.4 and operated in \emph{Mode~I}, i.e., one TCP connection was established for each TLS handshake.

In order to isolate network effects, CPU resources were constrained by pinning each container to a single logical CPU through explicit CPU affinity, so that endpoint compute capacity remained fixed across all measurements. The link capacity was controlled using Linux traffic control (\texttt{tc}) with a token bucket filter (TBF), sweeping the available bandwidth from \SI{1}{Mbps} to \SI{100}{Mbps}.

Because \emph{Mode~I} incurs per-handshake socket setup, teardown, and kernel I/O overheads that are not directly associated with cryptographic processing, we explicitly accounted for these effects in the endpoint parametrisation. In particular, for consistency with the first endpoint experiment using ML-KEM-1024, we introduced an additive I/O delay of $\delta_{\mathrm{client}} = \delta_{\mathrm{server}} = \SI{3}{ms}$,
which models the average time the CPU waits for socket-related I/O operations without performing handshakes or cryptographic computations.

\begin{figure}[ht]
    \centering
    \includegraphics[width=0.48\textwidth]{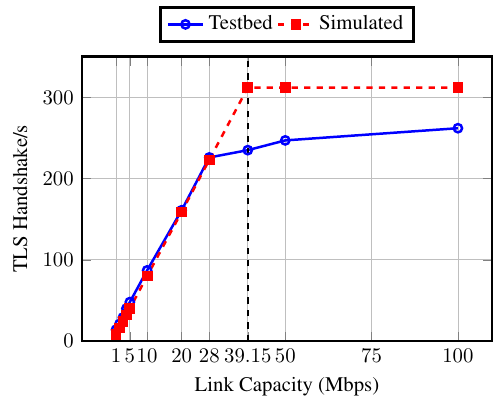}
    \caption{TLS throughput under bandwidth throttling.}
\label{fig:testbed-vs-sim}
\end{figure}

The maximum CPU-limited handshake rate (i.e., the endpoint-imposed plateau) was first estimated under the same single-CPU configuration used in the first endpoint experiment with ML-KEM-1024. In this network validation, the signature algorithm was additionally fixed to ML-DSA-87, and only the link capacity was varied. The resulting plateau occurs at approximately \SI{39.15}{Mbps}, at which the throughput reaches 312 handshakes per second and represents the asymptotic upper bound reached as the network ceases to be the bottleneck. This bandwidth therefore marks the transition between the bandwidth-limited (network-dominant) regime and the CPU-limited (endpoint-dominant) regime. As shown in Figure~\ref{fig:testbed-vs-sim}, the simulated throughput closely matches the testbed measurements across the full range of configured bandwidths, accurately capturing both the bandwidth-limited regime and the transition toward the CPU-limited plateau. In this experiment, the endpoint configuration matches that of Fig.~\ref{fig:ml-kem-exp1}, using ML-KEM-1024 for key exchange and ML-DSA-87 for authentication. The network byte budget is parameterised as \(B_{\mathrm{fix}}=700\), \(b_{\mathrm{pk}}=1568\), \(b_{\mathrm{ct}}=1568\), \(b_{\mathrm{cer}}=7219\), and \(b_{\mathrm{sig}}=4627\). Packet loss is set to $p=0$ since it is assumed to be negligible in the testbed and no artificial loss was introduced. The dashed line at \SI{39.15}{Mbps} in Figure~\ref{fig:testbed-vs-sim} delineates the two operational regimes: network-dominant constraints (left) and endpoint-dominant constraints (right). These results confirm that the proposed network service-time model provides a consistent mapping between link capacity and sustainable TLS handshake throughput, which is a key requirement for the calculator integration discussed in Section~\ref{sec:heur_calc}.

\section{Integration in network scenarios}
\label{sec:heur_calc}
The analytical model developed in the previous sections provides a tight upper bound on the sustainable handshake throughput for high-level security protocols, given the technical specifications of the endpoints and the capacity of the communication channel. In the simplest case of a single unicast connection between two endpoints, the achievable throughput is directly determined by the minimum service rate among the client, the server, and the network. This framework naturally extends to more general network scenarios, where two arbitrary endpoints are connected through a network graph with multiple nodes and one or more candidate paths. In this case, the relevant network quantity is no longer the capacity of a single link, but the \emph{end-to-end path capacity}, which depends on routing decisions and on the capacities of the links composing each path. Since the network service rate satisfies
\begin{equation}
    \mu_{\mathrm{net}}=\frac{C(1-p)}{8B},
\end{equation}
any reduction in the available path capacity $C$ directly translates into a lower sustainable handshake throughput.

A key feature of the model is that this relation can be inverted. Given a target handshake throughput between two endpoints (i.e., a \emph{user demand} expressed in handshakes per second), the network model yields the corresponding \emph{bandwidth demand}, namely the minimum end-to-end path capacity required to support that throughput under the assumed packet loss and handshake payload size. This inversion enables a direct mapping from application-level demands to network-level resource requirements.

\begin{figure}[t]
    \centering
    \includegraphics[width=0.35\textwidth]{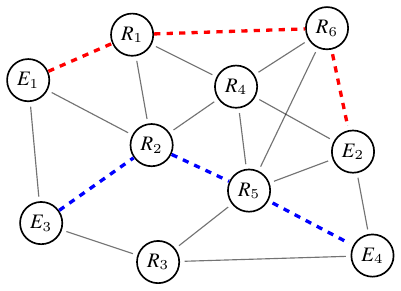}
    \caption{PQC calculator application to a simple mesh network topology.}
    \label{fig:network_topology}
\end{figure}

In a more sophisticated setting, such as the one implemented in the \texttt{telaQ}$^{\mathrm{TM}}$ simulation environment~\cite{telaQ}, this mapping allows user demands expressed in handshake rates to be translated into bandwidth demands on the network graph (Section~\ref{sec:pqc_calc}). The network itself is modeled as a set of nodes and links with fixed technical specifications and link capacities. Given multiple simultaneous unicast demands, the simulator autonomously solves a network optimization problem to determine feasible routing and resource allocation policies. In Figure~\ref{fig:network_topology}, two unicast connections between \(E_1\)–\(E_2\) and \(E_3\)–\(E_4\) generate handshake-rate demands that are translated into bandwidth requirements, and the simulator computes the corresponding feasible end-to-end paths under capacity constraints, which are highlighted.

In particular, the routing and capacity allocation problem can be formulated as a MCF optimisation problem with QoS constraints, where each commodity represents a unicast demand between a pair of endpoints. The objective may be, for example, to maximize the aggregate served demand, to maximize fairness, or to minimize congestion, subject to link capacity constraints and QoS requirements. The solution may involve single-path or multi-path routing, depending on the network configuration and optimization policy. Importantly, this optimisation is performed entirely by the network simulator and is independent of the cryptographic performance model.

Once the MCF optimisation produces a feasible allocation of end-to-end path capacities for each demand, the same analytical model can be applied in the forward direction to translate the allocated bandwidth back into the corresponding achievable handshake throughput for each connection. In this way, the calculator operates at the interface between application requirements and network resources: it maps user demands in terms of handshake rates into bandwidth demands prior to optimisation, and it maps the optimized bandwidth allocations back into achievable handshake performance after optimisation.

\subsection{Resource rescaling and endpoint sharing}
\label{sec:resource-rescaling}

When extending the model from a single unicast connection to multiple simultaneous unicast connections, a new issue arises: endpoint resources (logical CPUs, clock frequency, memory, network interface capacity) must be shared among multiple flows that terminate at the same endpoint. If the aggregate demand exceeds the physical capacity of the endpoint, the per-connection service rates predicted by the single-connection model are no longer simultaneously feasible and must be rescaled.

In this section, we introduce a simple and computationally efficient rescaling rule that enforces feasibility at each endpoint while providing a proportional and non-starving allocation of resources. Consider an endpoint $e$ (client or server) serving a set $\mathcal{K}_e$ of simultaneous unicast connections indexed by $k \in \mathcal{K}_e$.
Let $\lambda_{e,k}$ denote the requested handshake rate (handshakes per second) of connection $k$ served by $e$. Note that we can define the traffic associated with endpoint $e$ as the entire set of demands $\Lambda_e:=\{ \lambda_{e,k} \}_{k \in \mathcal{K}_e}$.

From the model in Section~\ref{sec:gen_end_model}, each handshake of connection $k$ induces an endpoint-side service time composed of a dominant asymmetric cryptographic component and an additive delay term. Let $c_{e,k}$ denote the total cryptographic cycle cost per handshake at endpoint $e$ for connection $k$ (i.e., the sum of the cycle costs of the modeled public-key operations, including the multiplicative overhead fraction $\alpha_e$). Note that this can be written as $c_{e,k}=\sum_{i \in \mathcal{O}_{e}^{k}} c_{e,i}^{k}$, where $\mathcal{O}_{e}^{k}$ is the set of public-key cryptographic operations performed by endpoint $e$ for connection $k$. 
Let $f_{e,k}$ denote the effective per-logical-CPU clock-equivalent  compute share (hereafter simply referred to as compute share) allocated by endpoint $e$ to connection $k$, with
\begin{equation}
f_{e,k}\ge 0,
\qquad
\sum_{k\in\mathcal K_e} f_{e,k}\le f_e.
\label{eq:clock_share_constraint}
\end{equation}
Then, using this allocation together with Eq.~\eqref{eq:time_operation} and Eq.~\eqref{eq:endpoint_service_time}, the per-handshake service time for connection $k$ at endpoint $e$ can be written as
\begin{equation}
T_{e,k}
=
(1+\alpha_e)\frac{c_{e,k}}{f_{e,k}\eta_e}
+\delta_{e}.
\label{eq:Tek_share_new}
\end{equation}
For a requested handshake rate $\lambda_{e,k}$, the corresponding compute share required to sustain that demand is obtained by solving
\begin{equation}
\lambda_{e,k}
=
\frac{\rho_e N_e}{
(1+\alpha_e)\dfrac{c_{e,k}}{f_{e,k}\eta_e}
+\delta_{e}},
\end{equation}
which yields
\begin{equation}
f^{\mathrm{req}}_{e,k}
= f_{e,k}^*:=
\frac{(1+\alpha_e)\lambda_{e,k}c_{e,k}}
{\eta_e\left(\rho_e N_e-\lambda_{e,k}\delta_{e}\right)}.
\label{eq:fek_req_new}
\end{equation}
Note that, more robustly, we set 
\begin{equation}
f^{\mathrm{req}}_{e,k}
=
\begin{cases}
f^{*}_{e,k}, & \text{if~} \rho_e N_e-\lambda_{e,k}\delta_{e}>0,\\
f_e, & \text{otherwise}.
\end{cases}
\end{equation}
If $\rho_e N_e-\lambda_{e,k}\delta_{e}\leq{0}$, we need to impose $f^{\mathrm{req}}_{e,k} = f_e$. In fact, this models the case when the requested demand is infeasible, so we try to allocate the maximum available compute share of endpoint $e$ to $k$. In case of competing demands, these will be rescaled according to the same rules as described below. In the worst-case scenario where all demands are infeasible, the resources will be equally shared among them.
The aggregate requested compute share at endpoint $e$ is therefore
\begin{equation}
f^{\mathrm{req}}_e=\sum_{k\in\mathcal K_e} f^{\mathrm{req}}_{e,k}.
\end{equation}

If $f^{\mathrm{req}}_e \le f_e$, the requested demands are feasible. Otherwise, the endpoint is oversubscribed and the requested shares must be proportionally rescaled. We therefore introduce the endpoint scaling factor
\begin{equation}
\gamma_e
=
\min\!\left(
1,\,
\frac{f_e}{f^{\mathrm{req}}_e}
\right),
\qquad
0 < \gamma_e \le 1 .
\end{equation}
The rescaled compute shares are then
\begin{equation}
f'_{e,k}
=
\gamma_e ~f^{\mathrm{req}}_{e,k}
=
\begin{cases}
f^{\mathrm{req}}_{e,k}, & \text{if ~} f^{\mathrm{req}}_e \le f_e,\\[6pt]
\dfrac{f_e}{f^{\mathrm{req}}_e}\, ~f^{\mathrm{req}}_{e,k}, & \text{if ~} f^{\mathrm{req}}_e > f_e,
\end{cases}
~~k\in\mathcal K_e .
\label{eq:rescaled_share_piecewise}
\end{equation}
By construction, the rescaled shares satisfy the endpoint total clock constraint
\begin{equation}
\sum_{k\in\mathcal K_e} f'_{e,k}\le f_e .
\end{equation}
The feasible endpoint-side handshake rate for connection $k$ is therefore
\begin{equation}
\lambda'_{e,k}
=
\frac{\rho_e N_e}{
(1+\alpha_e)\dfrac{c_{e,k}}{f'_{e,k}\eta_e}
+\delta_{e}
}.
\label{eq:lambda_rescaled_final}
\end{equation}
Therefore, after the rescaling step, one directly obtains the feasible endpoint-side handshake rates for each connection $k$ at the client and server, namely $\lambda'_{\mathrm{client},k}$ and $\lambda'_{\mathrm{server},k}$. 

\begin{figure}[b]
    \centering
    \includegraphics[width=0.35\textwidth]{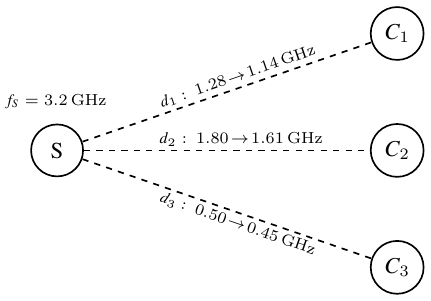}
    \caption{Proportional endpoint rescaling under CPU oversubscription.} \label{fig:clock_rescale}
\end{figure}

Figure~\ref{fig:clock_rescale} illustrates this mechanism for a simple server-clients topology, where multiple client demands $d_1,d_2,d_3$ compete for a fixed server clock budget $f_S$. In the illustrated example, the server consists of a single logical CPU operating at \(f_S=3.2\,\mathrm{GHz}\) and $\eta_S=1$. The aggregate requested compute across the three demands amounts to \(3.58\,\mathrm{GHz}\), exceeding the available clock budget. All demands are therefore proportionally scaled by the common reduction factor \(\gamma_S = 3.2/3.58 \approx 0.894\). The labels report both the requested and rescaled per-connection compute rates (in GHz), showing how proportional rescaling restores feasibility while preserving the relative demand ratios.

\subsection{PQC calculator}
\label{sec:pqc_calc}

The analytical components introduced in the previous sections can be aggregated into a functional module that we refer to as the \emph{PQC calculator}. The purpose of this module is to provide an interface between application-level handshake demands and network-level resource allocation. In particular, the calculator performs two main operations: (i) endpoint clock rescaling, which determines the maximum feasible demand supported by the endpoints, and (ii) demand--bandwidth conversion, which translates handshake-rate demands into bandwidth requirements and vice versa.

Consider the entire set $\mathcal{K}$ of unicast connections open in the network. Its index $k$ labels a generic connection in the set. Each connection has an associated demand represented by a requested handshake rate $\lambda_k$, expressed in handshakes per second. The total traffic is defined as the entire set of input demands $\Lambda:=\{ \lambda_k \}_{k \in \mathcal{K}}$. Each connection $k$ is associated with a pair of endpoints $e \in \{ \mathrm{client,\mathrm{server}}\}$, so we have $\lambda_{e,k}=\lambda_k$. Given the input total traffic $\Lambda$, each endpoint $e$ will have associated traffic $\Lambda_e$, corresponding to the set of connections served by that endpoint $e$.

For any given connection $k$, the first step consists of evaluating the feasibility of the requested rate $\lambda_k$  with respect to endpoint compute resources. This operation relies on the endpoint rescaling mechanism introduced in the previous section, which can be treated here as a black-box procedure. Applying the rescaling rule independently to the client, with traffic $\Lambda_{\mathrm{client}}$, and the server, with traffic $\Lambda_{\mathrm{server}}$, yields the feasible endpoint-side handshake rates $\lambda'_{\mathrm{client},k}$ and $\lambda'_{\mathrm{server},k}$ for connection $k$. These quantities represent the maximum feasible handshake rates that each endpoint can sustain for connection $k$ under the current resource allocation, capped by the requested demand $\lambda_k$, and therefore satisfy
$\lambda'_{\mathrm{client},k} \le \lambda_k$ and $\lambda'_{\mathrm{server},k} \le \lambda_k$. Since we cannot exceed the processing capability of either endpoint, the maximum rate supported by the endpoints for connection $k$ is determined by the bottleneck
\begin{equation}
\lambda^{\mathrm{ep}}_k =
\min\!\left(
\lambda'_{\mathrm{client},k},
\lambda'_{\mathrm{server},k}
\right).
\end{equation}

The next step converts this endpoint-feasible demand into a corresponding bandwidth requirement. Using the network service-rate relation in Eq.~(\ref{eq:mu_net}), the required bandwidth for connection $k$ can be obtained by inversion. Denoting by $B_k$ the number of handshake bytes associated with the selected protocol and primitive configuration, the requested bandwidth (RBW) becomes
\begin{equation}
C^{\mathrm{req}}_k =
\frac{8 B_k}{1-p}\,
\lambda^{\mathrm{ep}}_k .
\end{equation}

The entire set of bandwidth requests $\{C^{\mathrm{req}}_k\}_{k \in \mathcal{K}}$ constitutes the input to a network solver~\cite{telaQ}, which computes feasible end-to-end routes and allocates bandwidth across the network graph under link-capacity constraints. As a result of this optimization process, to each connection $k$ is assigned an allocated bandwidth denoted by $C^{\mathrm{alloc}}_k$ (with $C^{\mathrm{alloc}}_k \le C^{\mathrm{req}}_k$).

Once the network allocation has been determined, the calculator converts the allocated bandwidth back into the corresponding achievable handshake rate. Using again Eq.~(\ref{eq:mu_net}), the network-supported rate for connection $k$ is
\begin{equation}
\lambda'_{\mathrm{net},k} =
\frac{C^{\mathrm{alloc}}_k (1-p)}{8 B_k}.
\end{equation}
Since each step of the procedure can only reduce the achievable demand, the quantity $\lambda'_{\mathrm{net},k}$ represents the final satisfied handshake rate after accounting for endpoint limitations, routing constraints, and network capacity allocation. In other words, the calculator transforms the initial application demand into a feasible service level determined by the available computational and networking resources.

The quality of service (QoS) for connection $k$ is expressed by the ratio between satisfied and requested demand
\begin{equation}
\mathrm{QoS}_k =
\frac{\lambda'_{\mathrm{net},k}}{\lambda_k}.
\end{equation}
This quantity provides a simple performance indicator for network-planning scenarios, quantifying the fraction of the requested handshake workload that can be effectively served by the system under the given resource constraints.

In conclusion, the network solution will be provided by the entire set of network-supported rates $\{ \lambda'_{\mathrm{net,k}}\}_{k \in \mathcal{K}}$ or, equivalently, by the QoS set $\{ \mathrm{QoS}_k \}_{k \in \mathcal{K}}$.

\subsection{Strength and limitations}
This allocation rule provides \emph{proportional fairness in endpoint resource usage}. Each connection receives a share of the endpoint’s compute capacity that is proportional to its requested compute load. Connections using more expensive cryptographic primitives naturally consume more cycles per handshake and therefore obtain fewer handshakes per second for the same allocated share of compute.

Importantly, this rule does not enforce fairness in terms of handshake throughput across different primitives. Such fairness would require explicit weighting or normalization across heterogeneous per-handshake costs and is not the objective here. Instead, the goal is to fairly distribute the logical endpoint resource while guaranteeing graceful degradation: when the endpoint becomes overloaded, all connections are throttled proportionally and no connection is abruptly starved.

We deliberately perform proportional scaling in the resource domain (cycles or compute share capacity) rather than directly in the demand domain (handshakes per second). This avoids implicitly subsidizing computationally expensive primitives and preserves physical consistency with the endpoint capacity constraints.

The proposed rescaling rule is intentionally a \emph{zero-order local heuristic}. Each endpoint independently enforces its own capacity constraint by a single proportional projection step, without considering global coupling effects across the network or across multiple endpoints. This approach has several practical advantages: (i) it is computationally inexpensive and scales linearly with the number of incident connections per endpoint; (ii) it is stable and monotone, i.e., feasibility is always enforced and no flow is starved in a single step; (iii) it integrates naturally into network simulation workflows, where fast local corrections are preferred over expensive global optimization.

However, this method may not generally produce a globally optimal allocation. In networks with multiple shared endpoints and interacting bottlenecks, local rescaling may induce cascade effects: throttling at one endpoint modifies the load seen by another endpoint, which may in turn trigger further rescaling. The resulting allocation is feasible but not necessarily optimal with respect to any global objective (e.g., maximum aggregate throughput, max-min fairness, or utility maximization).

A principled global solution would require solving a joint optimization problem over all connections, subject to both network capacity constraints and endpoint compute constraints. Let $C_\ell$ denote the capacity of network link $\ell$, and let $\mathcal{K}_\ell$ denote the set of unicast connections whose selected path traverses link $\ell$. 
The resulting constraints can be written as
\begin{align}
\sum_{k \in \mathcal{K}_\ell} 8 B_k \lambda_k &\le C_\ell,
\quad \text{for each link } \ell,\\
\sum_{k \in \mathcal{K}_e} \lambda_k T_{e,k} &\le \rho_e N_e,
\quad \text{for each endpoint } e,
\end{align}
together with a chosen objective function (e.g., maximizing total served demand, max-min fairness, or proportional fairness). 
Depending on routing assumptions, this leads to large-scale linear, convex, or mixed-integer optimization problems, whose computational cost is incompatible with fast exploratory simulation and what-if analysis.

For this reason, the present work intentionally adopts the lightweight proportional rescaling rule as a practical approximation. It provides fast, stable, and interpretable feasibility enforcement while acknowledging that it does not achieve global optimality in highly coupled scenarios.

\section{Conclusion}
\label{sec:conclusion}
The results presented in this work demonstrate that it is possible to calculate a tight upper bound on the number of TLS and SSH handshakes per second achievable by a pair of endpoints in a quantum-safe network using PQC primitives. The presented model can be extended to other security protocols and integrated into a network simulator by leveraging the interfaces introduced in Section~\ref{sec:heur_calc}. The accuracy of the model depends on parameter estimation, which may be affected by noise, as observed for some primitives such as ML-DSA, where the variance of the mean CPU clock cycles is significant. The experimental data further suggest that the model is more accurate when cryptographic operation time is dominant, as expected, since this reduces the impact of confounding factors such as I/O delays.

\end{document}